\documentclass{aa}
\usepackage{graphicx}
\usepackage{txfonts}
\usepackage{natbib}
\bibpunct{(}{)}{;}{a}{}{,}
\begin{document}
%
   \title{Confirmation of the binary status of Cha H$\alpha$ 2 - \\a very young low-mass binary in Chamaeleon\thanks{Based on observations made with ESO telescopes at the Paranal Observatory under programme IDs 076.C-0292A, 076.C-0339B, 078.C-0535A, at the La Silla Observatory under programme ID 065.L-0144B, the Hubble Space Telescope under programme ID GO-8716 and on observations made with the European Southern Observatory telescopes obtained from the ESO/ST-ECF Science Archive Facility.}}
   \titlerunning{Confirmation of the binary status of Cha H$\alpha$ 2}

   \author{Tobias O.\,B. Schmidt
          \inst{1}
          \and
          Ralph Neuh\"auser\inst{1}
          \and
          Nikolaus Vogt\inst{2,3}
          \and
          Andreas Seifahrt\inst{1}
          \and
          Tristan Roell\inst{1}
          \and
          Ana Bedalov\inst{1}
          }

   \offprints{Tobias Schmidt, e-mail:~tobi@astro.uni-jena.de}

   \institute{Astrophysikalisches Institut und Universit\"ats-Sternwarte, Universit\"at Jena, Schillerg\"a\ss chen 2-3, 07745 Jena, Germany\\
              \email{tobi@astro.uni-jena.de}
         \and
             Departamento de F\'isica y Astronom\'ia, Universidad de Valpara\'iso, Avenida Gran Breta\~na 1111, Valpara\'iso, Chile
	 \and
	     Instituto de Astronom\'ia, Universidad Catolica del Norte, Avda.~Angamos 0610, Antofagasta, Chile
             }

   \date{Received 2007; accepted 2008}

 
  \abstract
   {Neuh\"auser \& Comer\'on (1998,\,1999) presented direct imaging evidence, as well as first spectra,
of several young stellar and sub-stellar M6- to M8-type objects in the Cha I dark cloud. One of these
objects is Cha H$\alpha$\,2, classified as brown dwarf candidate in several publications and
suggested as possible binary in Neuh\"auser et al. (2002).}
   {We have searched around Cha H$\alpha$\,2 for close and faint companions with adaptive optics imaging.}
   {Two epochs of direct imaging data were taken with the Very Large Telescope (VLT) Adaptive Optics instrument NACO in February
2006 and March 2007 in Ks-band together with a Hipparcos binary for astrometric calibration. Moreover, we took a J-band image in March 2007 to get color information. We retrieved an earlier image from 2005 from the European Southern Observatory (ESO) Science Archive Facility, increasing the available time coverage. After confirmation of common proper motion, we deduce physical parameters of the objects by spectroscopy, like temperature and mass.}
   {We find Cha H$\alpha$\,2 to be a very close binary of $\sim$\,0.16 arcsec separation, having a flux ratio of $\sim$\,0.91, thus having almost equal brightness and indistinguishable spectral
types within the errors. We show that the two tentative components of Cha H$\alpha$\,2 form a common proper
motion pair, and that neither component is a non-moving background object. We even find evidence for orbital motion.
A combined spectrum of both stars spanning optical and near-infrared parts of the spectral energy distribution
yields a temperature of 3000\,$\pm$\,100\,K, corresponding to a spectral type of M6\,$\pm$\,1 and a surface gravity of $\log{g}$\,=\,4.0$^{+0.75}_{-0.5}$, both from a comparison with GAIA model atmospheres. Furthermore, we obtained an optical extinction of A$_{\rm V}$\,$\simeq$ 4.3 mag from this comparison.}
   {We derive masses of $\sim$\,0.110\,M$_{\odot}$ ($\geq$\,\,0.070\,M$_{\odot}$) and $\sim$\,0.124\,M$_{\odot}$ ($\geq$\,\,0.077\,M$_{\odot}$) for the two components of Cha H$\alpha$ 2, i.e., probably low-mass stars, but one component could possibly be a brown dwarf.}


   \maketitle
%

\section{Introduction}

\citet{1998Sci...282...83N} found Cha H$\alpha$\,2, also called \object{ISO-ChaI 111}, to be a member of the Cha I star-forming cloud with an age of $\sim$\,2\,Myr. \citet{1999A&A...350..612N} then classified Cha H$\alpha$\,2 as a candidate brown dwarf of spectral type M6.5, using medium-resolution optical spectroscopy.
\citet{2000A&A...359..269C} presented evidence for near- to mid-infrared excess indicating a disk.
\citet{2001A&A...376L..22N} and \citet{2002ApJ...573L.115A} argued that the spectrum 
can be explained with an optically flat dust disk, but not with a flared disk, due to lack of
any prominent silicate feature previously expected for the disk of Cha H$\alpha$\,2.
\citet{2002A&A...384..999N} searched for companions around Cha H$\alpha$\,1\,-\,12
with the Hubble Space Telescope and presented the first indication for binarity of 
Cha H$\alpha$\,2 due to elongation of the Point Spread Function (PSF) in filters R and H$\alpha$.

The rotational period of 3.21\,$\pm$\,0.17\,days, found in photometric data 
by \citet{2003ApJ...594..971J}, confirmed the spectroscopic period 
of 2.9$^{+1.4}_{-1.0}$ days already proposed in \citet{2001A&A...379L...9J}. 
\citet{2004A&A...424..603N} observed nine stars in the Cha I dark cloud and 
found only at Cha H$\alpha$\,2 signs of accretion in H\,$\alpha$ and Pa\,$\beta$, 
making it the only object out of nine to show clear signs of accretion 
at a rate of $\sim$\,$10^{-10}$ M$_{\odot}$/yr and an inclination of $\sim$\,65 to 
75\,$^{\circ}$. Observations with XMM-Newton by \citet{2004A&A...423.1029S} 
revealed X-ray emission from Cha H$\alpha$\,2.
Recently, \citet{2005Sci...310..834A} succeeded in using new infrared spectra
to conclude that the disk around Cha H$\alpha$\,2 is still closer to
the flat disk model, even though a prominent enstatite silicate feature
could be found at 9.3\,$\mu$m.

Here, we present evidence for the binarity of Cha H$\alpha$\,2. In Sects.~2 \& 3, we present the observations, data reduction, and astrometric results. The available photometric properties of both components will be discussed in Sect.~4. The spectroscopy of the system is presented in Sect.~5. We end with conclusions in Sect.~6.


\section{Observations with VLT/NACO}

We observed Cha H$\alpha$\,2 in two epochs in February 2006 and in March 2007, see Table \ref{table:1} for the observations log. Included in Table \ref{table:1} are the data found in the archive from March 2005.

We observed with the European Southern Observatory (ESO) Very Large Telescope (VLT) instrument Naos-Conica \citep[NACO,][]{2003SPIE.4841..944L, 2003SPIE.4839..140R}. DIT, NDIT, NINT, and filter bands are listed in Table \ref{table:1}. In all cases, we used the S13 camera ($\sim$13 mas/pixel pixel scale) and the double-correlated read-out mode.

For the raw data reduction, we subtracted a mean dark from all the science frames and flatfield frames, then divided by the mean dark-subtracted flatfield and subtracted the mean background using \textit{eclipse\,/\,jitter}.

   \begin{figure}
   \resizebox{\hsize}{!}{\includegraphics{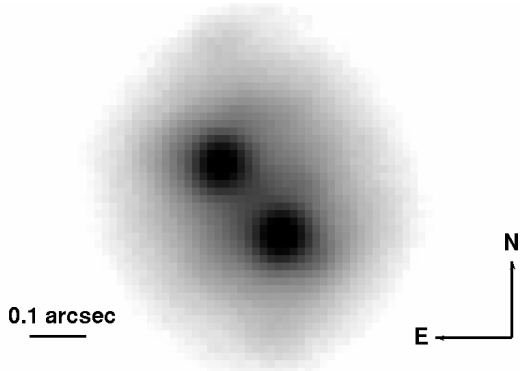}}
   \caption{NACO Ks-band image of the low-mass binary Cha H$\alpha$\,2 from 25 March 2005. The southwestern component appears slightly brighter, hence we call this component A (or SW), the fainter northeastern component B (or NE).}
   \label{image}
   \end{figure}

\begin{table}
\caption{VLT/NACO observations log}
\label{table:1}
\begin{tabular}{lcccccccc}
\hline\hline
JD - 2453400 & Date of     & DIT & NDIT & No.~(a) & Filter\\
$[\mathrm{days}]$     & observation & [s] &      & images  &       \\
\hline
\ \ 54.62360\,(b) & 25 Mar 2005 & 30 & 2 & 23 & Ks \\
   382.67884 & 16 Feb 2006      & 10 & 6 & 20 & Ks \\
   760.70760 &  1 Mar 2007      & 60 & 2 & 15 & Ks \\
   761.66193 &  2 Mar 2007      & 30 & 2 & 24 & J  \\
\hline
\end{tabular}
Remarks: (a) Each image consists of the number of exposures given in column 4 times the individual integration time given in column 3. (b) Retrieved from the ESO Science Archive Facility.
\end{table}

\begin{table}
\caption{Proper motions of Cha H$\alpha$\,2}
\label{table:2}
\begin{center}
\begin{tabular}{lr@{\,$\pm$\,}lr@{\,$\pm$\,}l}
\hline\hline
Source &  \multicolumn{2}{c}{PM RA}    & \multicolumn{2}{c}{PM DEC}  \\
       &  \multicolumn{2}{c}{[mas/yr]} & \multicolumn{2}{c}{[mas/yr]}\\
\hline
USNO-B1 (a) & -20  &17   & -14  &4   \\
PSSPMC (b)  & -23  &17   &  +1  &17  \\
SSS-FORS1   & -19.4&14.9 & -0.3 &14.9\\
SSS-SofI    & -23.0&14.1 & +7.8 &14.1\\
Mean        & -21.8&8.8  & +3.2 &8.8 \\
\hline
\end{tabular}
\end{center}
Remarks: (a) This value \citep{2003AJ....125..984M} was not used in the 
final proper motion due to its unreasonably small error of PM in declination,
which makes it inconsistent with the measurements using the SuperCOSMOS Sky Survey in comparison to data from the New Technology Telescope (NTT) instrument Son of ISAAC (SSS-SofI). (b) From the Pre-main Sequence Stars Proper Motion Catalog \citep{2005A&A...438..769D}.
\end{table}

In all three images, Cha H$\alpha$\,2 is clearly resolved in a double object with an apparent separation of $\sim$160\,mas at a
position angle of $\sim$\,40$^{\circ}$, see Fig.~\ref{image}. The southwestern component appears slightly brighter, hence we call 
this component A (or SW), the fainter northeastern component B (or NE).

\section{Astrometry}

\begin{table*}
\caption{Astrometric calibration and astrometric results for Cha H$\alpha$\,2 AB }
\label{table:3}
\centering
\begin{tabular}{l|ccc|l|ccc}
\hline\hline
JD - 2453400 & Calibration & pixel scale   & orientation & JD - 2453400 & Target & separation  & PA (a) \\
$[\mathrm{days}]$  & binary & [mas/pixel] & [$\deg$] & [days] & & [mas] &  [$\deg$]  \\
\hline
\ \ 54.76919 & HIP 73111 & 13.22 $\pm$ 0.28 & -0.17 $\pm$ 1.84 & \ \ 54.62360 & Cha H$\alpha$\,2 AB &165.7 $\pm$ 3.7 & 40.16 $\pm$ 1.84\\
388.84537 & HIP 73357 & 13.24 $\pm$ 0.07 &  0.18 $\pm$ 0.46 & 382.67884 & Cha H$\alpha$\,2 AB &167.0 $\pm$ 1.4 & 41.01  $\pm$ 0.47\\
760.76369 & HIP 73357 & 13.24 $\pm$ 0.07 &  0.34 $\pm$ 0.46 & 760.70760 & Cha H$\alpha$\,2 AB &158.0 $\pm$ 1.8 & 39.73  $\pm$ 0.50\\
\hline
\end{tabular}
\begin{flushleft}
Remarks: All Ks-band images. (a) PA is measured from N over E to S. 180 $\deg$ to be added to position angle (PA) if seen from fainter component.
\end{flushleft}
\end{table*}

   \begin{figure}
   \resizebox{\hsize}{!}{\includegraphics{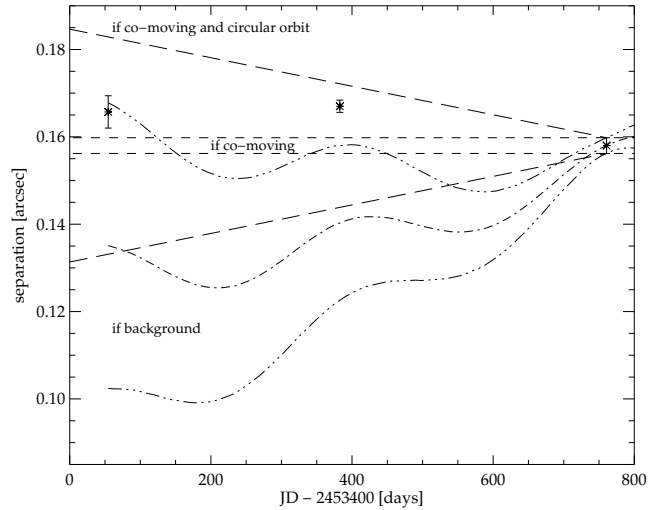}}
   \caption{Observed separation between the two components of Cha H$\alpha$\,2. The archival data point from 2005 and our two
measurements from 2006 and 2007 are shown. The short-dashed lines enclose the area for constant separation. The dash-dotted line is the change expected if the B component is a non-moving background star. The opening cone enclosed by the dash-dotted lines with more dots are its estimated errors. The waves of this cone show the differential parallactic motion, which has to be taken into account if the other component is a non-moving background star.
The opening long-dashed cone
indicates the amplitude of possible orbital motion.}
   \label{Sep1}
   \end{figure}

   \begin{figure}
   \resizebox{\hsize}{!}{\includegraphics{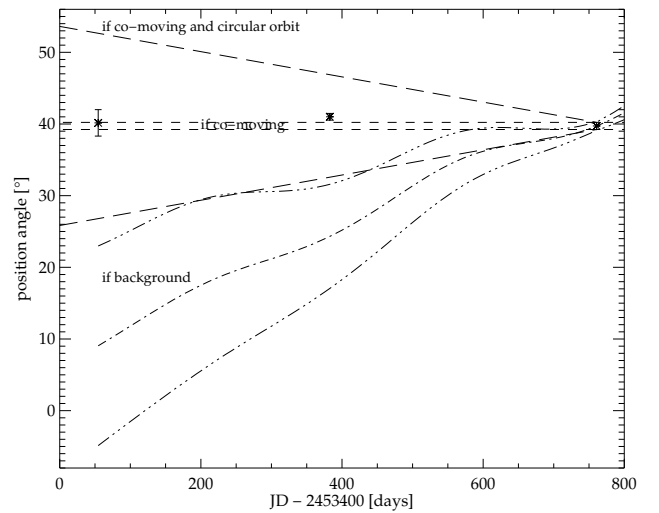}}
   \caption{Observed position angle between the two components of Cha H$\alpha$\,2 (measured from the brighter component north over east to south). See Fig. \ref{Sep1} for details.}
   \label{Pos1}
   \end{figure}

To check for common proper motion of the two tentative components of Cha H$\alpha$\,2, we used the proper motion (here PM) of Cha H$\alpha$\,2 published in the literature (USNO-B and PSSPMC); we calculated additional values by comparing SuperCOSMOS Sky Survey (here SSS) data from 1985 with images from VLT/FORS1 (epoch 1999, published by Comeron et al. 2000, obtained directly from F. Comer\'on, priv. com.)
and New Technology Telescope instrument Son of ISAAC (NTT/SofI) from 2000, observed by us (10\,$\times$\,50\,$\times$\,1.2 second integrations), reduced in the same way as the NACO data with eclipse. See Table \ref{table:2} for the results (originally from \citet{2005Gaedke}, but revised here by us).
We use the mean proper motion for checking, whether the two objects show common proper motion below.

We calibrated the NACO data using the wide binary star HIP 73357 for our two measurements in 2006 and 2007,
and the closer binary HIP 73111 for which we found archived data taken in the same night as the Cha H$\alpha$\,2 observations in 2005, resulting in the astrometric calibration summarized in Table \ref{table:3}. The error bars of pixel scale and orientation include, possible orbital motion of the respective binary since its measurement
by Hipparcos -- i.\,e., a maximum change in separation due to possible orbital motion for circular edge-on orbit and a maximum change in position angle for circular pole-on orbit -- as well as the uncertainties in parallax, total mass, position measurements in our images and separation at the epoch of Hipparcos of the respective binary system.

To determine the positions of both components we constructed a reference PSF from the binary itself by using the undeformed 
remote sides of both objects shifted to the same peak position. Thus, we obtained a clean reference PSF for each single image.
With IDL/starfinder, we scaled and shifted the reference PSF simultaneously to both components in each of our individual images by minimizing the residuals. As a result, we obtained positions (see Table~\ref{table:3}) and relative photometry (see Table~\ref{table:4}), including realistic error estimates for each object by averaging the results of all single images taken within each epoch.

The development of separation and position angle with time is shown in Figs.~\ref{Sep1}, \ref{Pos1}, \ref{Sep2}, and \ref{Pos2}.
We can exclude by 0.9\,$\sigma$ \& 1.5\,$\sigma$ (Fig. 2); 2.2\,$\sigma$ \& 2.3\,$\sigma$ (Fig. 3); 1.4\,$\sigma$ \& 1.0\,$\sigma$ (Fig. 4); 2.1\,$\sigma$ \& 1.9\,$\sigma$ (Fig. 5)
that either one of the tentative components of Cha H$\alpha$\,2 is a non-moving background object.
Viz, for both components the background hypothesis is rejected by $\geq$\,3.7\,$\sigma$, respectively.
For two objects slightly above the sub-stellar mass limit (see below) and the given separation (at $\sim168$ pc),
the orbital period is $\sim 110$ yrs, so that the maximum change in separation due to orbital
motion (for circular edge-on orbit) is $\sim 12$ mas/yr ($\sim 3\,^{\circ}$/yr in PA for pole-on orbit).
While orbital motion could be detected in PA only from one point in 2006 in contrast to 2007, with 1.9\,$\sigma$ significance, we detect orbital motion as deviation of the separation in 2005 and 2006 in contrast to the 2007 measurement, with {1.9\,$\sigma$ \& 3.9\,$\sigma$ significance. From the negligible change in PA during our $\sim$\,2 years of epoch difference, we can exclude a circular pole on orbit by 2.8\,$\sigma$, however, we cannot rule out a highly eccentric pole-on orbit.

In Fig.~\ref{Hubble}, the data from Hubble Space Telescope by \citet{2002A&A...384..999N} are also included. Only the plot for the assumption of component B being a non-moving background star is shown here. The Hubble Space Telescope data hence yield another 2.2\,$\sigma$ \& 2.3\,$\sigma$ for component A \& B not being a non-moving background star, respectively.

   \begin{figure}
   \resizebox{\hsize}{!}{\includegraphics{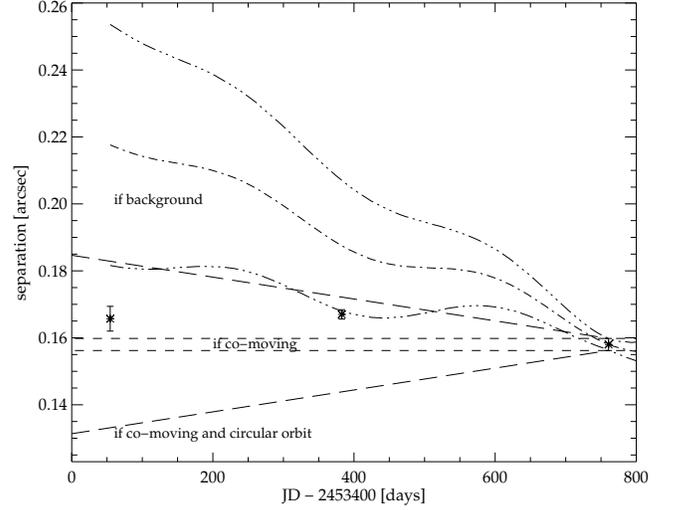}}
   \caption{Observed separation between the two components of Cha H$\alpha$\,2. Same as  Fig.~\ref{Sep1}, but the opening cone enclosed by the dashed-dotted lines is the change expected if the A component is a non-moving background star.}
   \label{Sep2}
   \end{figure}

   \begin{figure}
   \resizebox{\hsize}{!}{\includegraphics{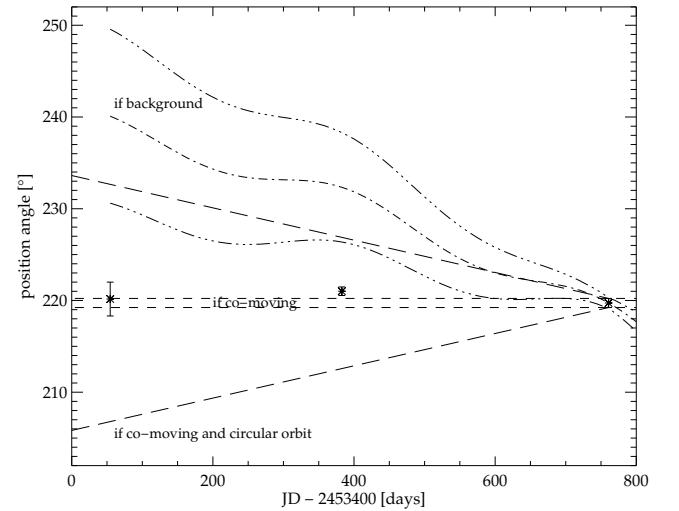}}
   \caption{Same as Fig.~\ref{Pos1}, but position angle between the two components measured from the B component
and assuming the A component is a non-moving background star.}
   \label{Pos2}
   \end{figure}

   \begin{figure*}
   \resizebox{\hsize}{!}{\includegraphics[angle=90]{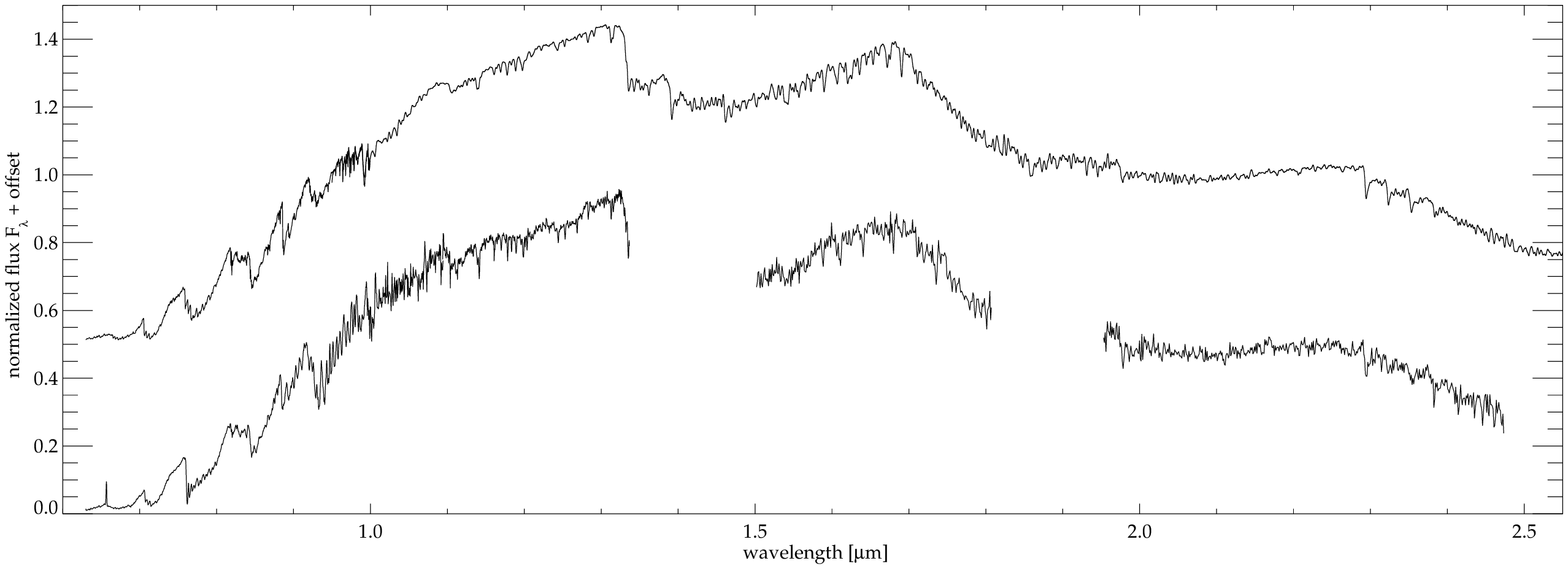}}
   \caption{Bottom: Spectrum of both components from the optical part to the near-infrared combined from data from \citet{2000A&A...359..269C} and data from the ESO Paranal and La Silla instruments ISAAC and SofI retrieved from the ESO Science Archive Facility. Top: The best fit to the spectrum:~GAIA models computed with the PHOENIX code by \citet{2005ESASP.576..565B} for a temperature of 3000\,K, $\log{g}$ of 4.0 and a visual extinction A$_{\rm V}\simeq 4.3$ mag. Note that the missing
flux depression in J-band (between 1.2 and 1.3 micron) as well as in the peak of the H-band (around 1.65 micron) is due to missing FeH opacities in the spectral models.
Note the strong H$\alpha$ emission in Cha H$\alpha$\,2 at 0.6568 $\mu$m.}
   \label{SpecAll}
   \end{figure*}

While the difference seen in PA and the separation between the different observations are consistent with common proper motion and some orbital motion, a possible difference in proper motion between both objects of up to a few mas/yr cannot be excluded from the data. Such a difference in proper motion is typical for the velocity dispersion in star forming regions like Cha I \citep{2005A&A...438..769D}, so that we cannot yet exclude that both objects are independent members of Cha I, but not orbiting each other. However, it is extremely unlikely that two objects with almost the same temperatures, J \& K magnitudes, gravities, radii, proper motions, and extinctions are located within less than 170 mas. Even if this would be the case, age, and distance would be the same as assumed below, hence also the mass estimation.

\section{Photometry}

As described in the last section, we obtained from the PSF fitting of both components also the relative flux ratio, see Table \ref{table:4}. We considered these flux ratios in the astrometric proper motion analysis since the background
hypotheses is based on proper motions obtained by centroid measurements of the unresolved system. Hence, if one
component were a non-moving background object, the centroid (and thus the proper motion of the combined object)
would only shift by about half the relative change in separation between both objects.

Using the unresolved photometry of Cha H$\alpha$\,2 from the Two Micron All Sky Survey (2MASS) catalogue of $J$\,=\,12.210\,$\pm$\,0.024\,mag and $K$\,=\,10.675\,$\pm$\,0.021\,mag, we obtain the photometry of each component using 
our measured flux ratios, see Table \ref{table:5}. We see no evidence for photometric variability.

   \begin{figure}
   \resizebox{\hsize}{!}{\includegraphics{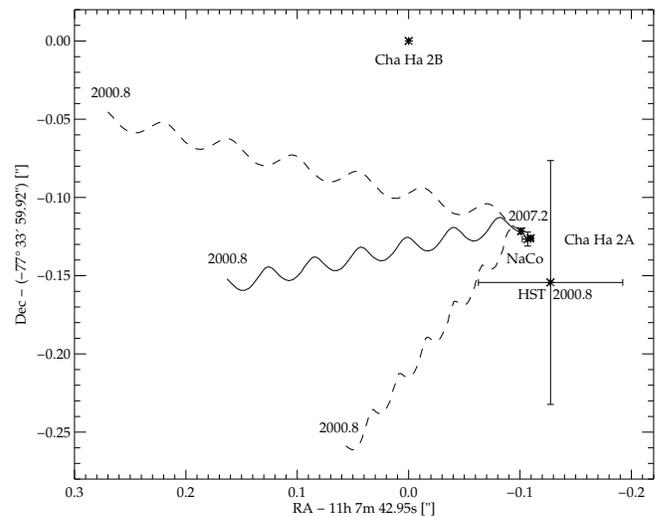}}
   \caption{Plot of the true and expected position of Cha H$\alpha$\,2 A if Cha H$\alpha$\,2 B is a non-moving background star. Shown are the NACO measurements from 2005 to 2007, as in Figs.~\ref{Sep1}\,-\,\ref{Pos2}, and, in addition, the measurement from 2000 using the Hubble Space Telescope in R-band by \citet{2002A&A...384..999N}. As in Fig.~\ref{Sep1}, the line showing parallactic wobbles is the expected position of component A if component B is a non-moving background star, in this case from epoch 2000.8 until 2007.2, from left to right. The opening dashed cone are its errors. The HST data point obtained at epoch 2000.8 is significantly different (2.3\,$\sigma$) from the expectation, if background.}
   \label{Hubble}
   \end{figure}

\section{Spectroscopy}

We combined all available spectra of Cha H$\alpha$\,2 from the archive and an optical spectrum
obtained directly from F. Comer\'on (priv. com.) published in \citet{2000A&A...359..269C} 
to create a combined spectrum with a nearly complete coverage from 0.64 $\mu$m to 2.45 $\mu$m.
The data were reduced in a standard manner, doing flat field correction, sky subtraction, spectrum
extraction, wavelength calibration, and standard correction with G stars. The spectra are not flux-calibrated. We found the relative offsets from the overlaps of the spectra.

We fitted this spectrum with GAIA COND models computed with the PHOENIX code by \citet{2005ESASP.576..565B}.
COND represents the case where dust forms, as in the similar DUSTY models, but rains out completely from the photosphere into deeper layers and, hence, the region where $\tau$\,=\,1 is free of dust. Since little or no dust forms at temperatures above $\sim$\,2400\,K both models with different treatment of dust formation give the same result at the given temperatures.
Our best fit results give an effective temperature of 3000\,$\pm$\,100\,K,
a surface gravity $\log{g} $\,=\,4.0$^{+0.75}_{-0.5}$
and a visual extinction of A$_{\rm V}$\,=\,4.3\,$\pm$\,0.25\,mag,
see Fig.~\ref{SpecAll} and the online material for a color version.
We note that our extinction value is in between the A$_{\rm V}$ of 7.9\,mag measured in a slope in the A$_{\rm V}$ map from \citet{2006A&A...447..597K} and the first ever measured
value of 3.55\,mag by \citet{1998Sci...282...83N}.
Given the quality of our fit, we find here further evidence for the physical binarity of Cha H$\alpha$\,2 being a double of almost identical mid- to late- M-type objects. 
If one of the components would be a background (or foreground) object, both the
spectral type and the extinction of the components would differ and the unresolved spectrum of the binary would not
be consistent with one spectral type over the whole spectral range.

The derived temperature can be converted to a spectral type of M6\,$\pm$\,1 with the empirical temperature scale for young M-type objects from \citet{1999ApJ...525..466L}. Moreover, our combined spectrum in Fig.~\ref{SpecAll} resembles very much the spectrum of the M5.75 object $\epsilon$ Cha 10 in \citet{2004ApJ...616.1033L}.

We also obtained low-S/N ($\sim$ 15) K-band spectra of both objects with VLT/NACO, see Fig.~\ref{NaCoSpec}. They are indistinguishable, giving further evidence that neither of the two objects are background or foreground stars, which would exhibit a significantly different K-band spectrum.

   \begin{figure}
   \resizebox{\hsize}{!}{\includegraphics[angle=90]{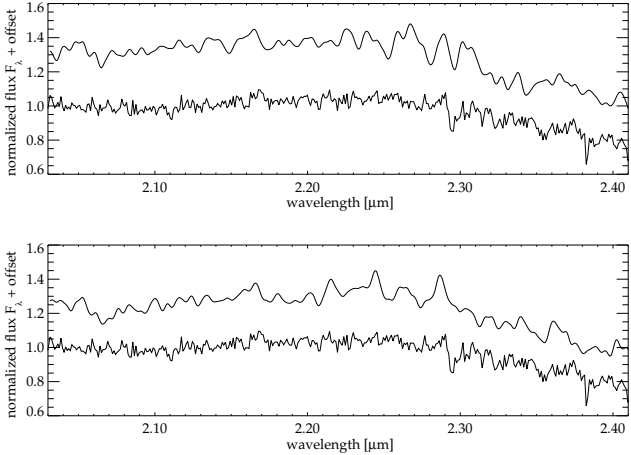}}
   \caption{Observed K-band NACO spectra of the Cha H$\alpha$\,2 system. In each panel, one of the components of the system is shown (top in each panel) compared to the higher S/N spectrum of both objects combined from Fig.~\ref{SpecAll} (bottom in each panel). Top: Component A (southwestern component). Bottom: Component B (northeastern component). Both spectra are dereddened using the visual extinction of A$_{\rm V}\simeq 4.3$ mag derived from the spectra shown in Fig.~\ref{SpecAll}.}
   \label{NaCoSpec}
   \end{figure}

\begin{table}
\caption{Flux ratios of the Cha H$\alpha$\,2 system}
\label{table:4}
\centering
\begin{tabular}{lccc}
\hline\hline
JD - 2453400 [days] & Filter & B/A flux ratio \\
\hline
\ \ 54.62360 & Ks & 0.931 $\pm$ 0.024\\
   382.67884 & Ks & 0.908 $\pm$ 0.014\\
   760.70760 & Ks & 0.894 $\pm$ 0.053\\
   Mean      & Ks & 0.910 $\pm$ 0.019\\
\hline
   761.66193 & J  & 0.893 $\pm$ 0.040\\
\hline
\end{tabular}
\end{table}

\begin{table}
\caption{Brightnesses of the Cha H$\alpha$\,2 system}
\label{table:5}
\centering
\begin{tabular}{lcccc}
\hline\hline
Component & J-band & Ks-band \\
\hline
Cha H$\alpha$\,2 A (SW) & 12.903 $\pm$ 0.033 & 11.378 $\pm$ 0.024\\
Cha H$\alpha$\,2 B (NE) & 13.026 $\pm$ 0.035 & 11.480 $\pm$ 0.024\\
\hline
\end{tabular}
\end{table}

\section{Conclusions}

From the photometry, the visual extinction, an extinction law to convert the visual to a near-infrared extinction by \citet{1985ApJ...288..618R}, a bolometric correction BC$_{\rm K}$ of 3.0\,$\pm$\,0.1\,mag from \citet{2004AJ....127.3516G} and the distance to a particular member of the Cha I cloud of 168$^{+28}_{-24}$ from \citet{1999A&A...352..574B} (calculated by HIPPARCOS measurements of 4 members of Cha I and resulting in an average distance of 168$^{+14}_{-12}$ to the cloud), we find luminosities log$(L/L_{\odot})$ of $-1.21\pm0.14$ \& $-1.25\pm0.14$, and radii of 0.92$_{\,-0.14}^{\,+0.19}$ \& 0.88$_{\,-0.13}^{\,+0.18}$ R$_{\odot}$ for Cha H$\alpha$\,2 A (SW) and B (NE), respectively.
This makes both components more massive than the more massive component of the eclipsing binary brown dwarf found by \citet{2006Natur.440..311S} in Orion of M\,=\,0.054\,$\pm$\,0.005\,M$_{\odot}$ (assuming similar age, which is justified).

It is, a priori, not known to which of both components the photometric
rotation period refers. However, since both stars are very young and have
been formed in similar environmental conditions, we may assume that they
rotate at the same speed. In this case, we can conclude from the photometric
period of 3.21\,$\pm$\,0.17\,days \citep{2003ApJ...594..971J} and the
v~sin\,i of 12.8\,$\pm$\,1.2\,km/s \citep{2001A&A...379L...9J}
that Cha H$\alpha$\,2 A and B have inclinations of
36.1 to 143.9\,$\deg$ and 40.4 to 139.6\,$\deg$, respectively,
consistent with $\sim$\,65 to 75\,$\deg$ by \cite{2004A&A...424..603N}.
Assuming an alignment of the orbital axis and the rotational axis of both components,
this is also consistent with our finding of astrometric orbital motion in the form of a change in separation rather than in a change of the PA, i.e., with an orbital plane being more edge-on than pole-on.

Given the luminosities, temperatures, radii, and gravities, the masses
of both components are 0.308$^{\,+1.430}_{\,-0.231}$\,M$_{\odot}$ and 0.281$^{\,+1,302}_{\,-0.210}$\,M$_{\odot}$ for A and B,
thus leaving Cha H$\alpha$\,2 AB above the stellar -- brown dwarf boundary.
However, component B could still be a brown dwarf, according to this error estimation.
Although we arrive at masses that could possibly be higher than the sun's (given our large error bars), this can be ruled out due to the fact that such very young objects could be of approximately same size as the sun, but would not have a similar surface gravity at the same time.
Using theoretical models by \citet{1998A&A...337..403B}, we find a best fit for masses of 0.124\,M$_{\odot}$ and 0.110\,M$_{\odot}$ for Cha H$\alpha$ 2 A and B for an age of 2 Myr, a temperature of $\sim$\,3025\,K, a $\log{g}$ of $\sim$\,3.6, and the luminosity values as derived above of log$(L/L_{\odot})$ of -1.21 \& -1.25, but we would like to mention that the determination of ages and masses from evolutionary models is uncertain until at least 10 Myrs \citep{2005astro.ph..9798C}. These mass values are in good agreement with
the mass for the unresolved Cha H$\alpha$ 2 of 0.14\,M$_{\odot}$ from \citet{2005ApJ...626..498M}.
While there are currently almost one hundred systems known of even lower total mass \citep{2007lyot.confR..45S}, a review of the field and the existence of very low mass binaries at different ages can be found in \citet{2007prpl.conf..427B}.

After submission of our manuscript \citet{2007arXiv0708.3851A} reported the possible binarity of Cha H$\alpha$ 2 found in their data from 2005, included as 1st epoch in our common proper motion analysis. We note that separation, position angle, and Ks-band magnitudes measured coincide well within the 1\,$\sigma$ error bars of each other. \citet{2007arXiv0708.3851A} could not show common proper motion because they had only one epoch available.

The projected separation between A and B corresponds to $\sim 27$ AU at $\sim 168$ pc,
so that further investigation of the interaction between the two components in
Cha H$\alpha$ 2 and its disk (with silicate feature) would be very important.

\begin{acknowledgements}
TOBS acknowledges support from Evangelisches Studienwerk e.V. Villigst.

Moreover he would like to thank F. Comer\'on for providing FORS1 images as well as an electronic optical spectrum of Cha H$\alpha$\,2 from his publication \citet{2000A&A...359..269C}, P. Hauschildt for the electronic GAIA models, J. Kainulainen for the electronic extinction map of the Cha I cloud for accurate reading of the extinction values, an anonymous referee for helpful comments and Amy Mednick for language editing.

NV acknowledges support by FONDECYT grant 1061199.

AB and RN would like to thank DFG for financial support in projects NE 515 / 13-1 and 13-2.

This publication makes use of data products from the Two Micron All Sky Survey, which is a joint project of the University of Massachusetts and the Infrared Processing and Analysis Center/California Institute of Technology, funded by the National Aeronautics and Space Administration and the National Science Foundation.

We use imaging data from the SuperCOSMOS Sky Survey, prepared and hosted by Wide Field Astronomy Unit, Institute for Astronomy, University of Edinburgh, which is funded by the UK Particle Physics and Astronomy Research Council.

This research has made use of the VizieR catalogue access tool and the Simbad database, both operated at the Observatoire Strasbourg.
\end{acknowledgements}

\bibliographystyle{aa}

\begin{thebibliography}{30}
\expandafter\ifx\csname natexlab\endcsname\relax\def\natexlab#1{#1}\fi

\bibitem[{{Ahmic} {et~al.}(2007){Ahmic}, {Jayawardhana}, {Brandeker}, {Scholz},
  {van Kerkwijk}, {Delgado-Donate}, \& {Froebrich}}]{2007arXiv0708.3851A}
{Ahmic}, M., {Jayawardhana}, R., {Brandeker}, A., {et~al.} 2007, ArXiv
  e-prints, 07083851

\bibitem[{{Apai} {et~al.}(2005){Apai}, {Pascucci}, {Bouwman}, {Natta},
  {Henning}, \& {Dullemond}}]{2005Sci...310..834A}
{Apai}, D., {Pascucci}, I., {Bouwman}, J., {et~al.} 2005, Science, 310, 834

\bibitem[{{Apai} {et~al.}(2002){Apai}, {Pascucci}, {Henning}, {Sterzik},
  {Klein}, {Semenov}, {G{\"u}nther}, \& {Stecklum}}]{2002ApJ...573L.115A}
{Apai}, D., {Pascucci}, I., {Henning}, T., {et~al.} 2002, \apjl, 573, L115

\bibitem[{{Baraffe} {et~al.}(1998){Baraffe}, {Chabrier}, {Allard}, \&
  {Hauschildt}}]{1998A&A...337..403B}
{Baraffe}, I., {Chabrier}, G., {Allard}, F., \& {Hauschildt}, P.~H. 1998, \aap,
  337, 403

\bibitem[{{Bertout} {et~al.}(1999){Bertout}, {Robichon}, \&
  {Arenou}}]{1999A&A...352..574B}
{Bertout}, C., {Robichon}, N., \& {Arenou}, F. 1999, \aap, 352, 574

\bibitem[{{Brott} \& {Hauschildt}(2005)}]{2005ESASP.576..565B}
{Brott}, I. \& {Hauschildt}, P.~H. 2005, in ESA Special Publication, Vol. 576,
  ESA Special Publication, ed. C.~{Turon}, K.~S. {O'Flaherty}, \& M.~A.~C.
  {Perryman}, 565

\bibitem[{{Burgasser} {et~al.}(2007){Burgasser}, {Reid}, {Siegler}, {Close},
  {Allen}, {Lowrance}, \& {Gizis}}]{2007prpl.conf..427B}
{Burgasser}, A.~J., {Reid}, I.~N., {Siegler}, N., {et~al.} 2007, in Protostars
  and Planets V, ed. B.~{Reipurth}, D.~{Jewitt}, \& K.~{Keil}, 427--441

\bibitem[{{Chabrier} {et~al.}(2005){Chabrier}, {Baraffe}, {Allard}, \&
  {Hauschildt}}]{2005astro.ph..9798C}
{Chabrier}, G., {Baraffe}, I., {Allard}, F., \& {Hauschildt}, P.~H. 2005, ArXiv
  Astrophysics e-prints

\bibitem[{{Comer{\'o}n} {et~al.}(2000){Comer{\'o}n}, {Neuh{\"a}user}, \&
  {Kaas}}]{2000A&A...359..269C}
{Comer{\'o}n}, F., {Neuh{\"a}user}, R., \& {Kaas}, A.~A. 2000, \aap, 359, 269

\bibitem[{{Ducourant} {et~al.}(2005){Ducourant}, {Teixeira}, {P{\'e}ri{\'e}},
  {Lecampion}, {Guibert}, \& {Sartori}}]{2005A&A...438..769D}
{Ducourant}, C., {Teixeira}, R., {P{\'e}ri{\'e}}, J.~P., {et~al.} 2005, \aap,
  438, 769

\bibitem[{{Gaedke}(2005)}]{2005Gaedke}
{Gaedke}, A. 2005, Diploma thesis, FSU Jena

\bibitem[{{Golimowski} {et~al.}(2004){Golimowski}, {Leggett}, {Marley}, {Fan},
  {Geballe}, {Knapp}, {Vrba}, {Henden}, {Luginbuhl}, {Guetter}, {Munn},
  {Canzian}, {Zheng}, {Tsvetanov}, {Chiu}, {Glazebrook}, {Hoversten},
  {Schneider}, \& {Brinkmann}}]{2004AJ....127.3516G}
{Golimowski}, D.~A., {Leggett}, S.~K., {Marley}, M.~S., {et~al.} 2004, \aj,
  127, 3516

\bibitem[{{Joergens} {et~al.}(2003){Joergens}, {Fern{\'a}ndez}, {Carpenter}, \&
  {Neuh{\"a}user}}]{2003ApJ...594..971J}
{Joergens}, V., {Fern{\'a}ndez}, M., {Carpenter}, J.~M., \& {Neuh{\"a}user}, R.
  2003, \apj, 594, 971

\bibitem[{{Joergens} \& {Guenther}(2001)}]{2001A&A...379L...9J}
{Joergens}, V. \& {Guenther}, E. 2001, \aap, 379, L9

\bibitem[{{Kainulainen} {et~al.}(2006){Kainulainen}, {Lehtinen}, \&
  {Harju}}]{2006A&A...447..597K}
{Kainulainen}, J., {Lehtinen}, K., \& {Harju}, J. 2006, \aap, 447, 597

\bibitem[{{Lenzen} {et~al.}(2003){Lenzen}, {Hartung}, {Brandner}, {Finger},
  {Hubin}, {Lacombe}, {Lagrange}, {Lehnert}, {Moorwood}, \&
  {Mouillet}}]{2003SPIE.4841..944L}
{Lenzen}, R., {Hartung}, M., {Brandner}, W., {et~al.} 2003, in Presented at the
  Society of Photo-Optical Instrumentation Engineers (SPIE) Conference, Vol.
  4841, Instrument Design and Performance for Optical/Infrared Ground-based
  Telescopes. Edited by Iye, Masanori; Moorwood, Alan F. M. Proceedings of the
  SPIE, Volume 4841, pp. 944-952 (2003)., ed. M.~{Iye} \& A.~F.~M. {Moorwood},
  944--952

\bibitem[{{Luhman}(1999)}]{1999ApJ...525..466L}
{Luhman}, K.~L. 1999, \apj, 525, 466

\bibitem[{{Luhman}(2004)}]{2004ApJ...616.1033L}
{Luhman}, K.~L. 2004, \apj, 616, 1033

\bibitem[{{Mohanty} {et~al.}(2005){Mohanty}, {Jayawardhana}, \&
  {Basri}}]{2005ApJ...626..498M}
{Mohanty}, S., {Jayawardhana}, R., \& {Basri}, G. 2005, \apj, 626, 498

\bibitem[{{Monet} {et~al.}(2003){Monet}, {Levine}, {Canzian}, {Ables}, {Bird},
  {Dahn}, {Guetter}, {Harris}, {Henden}, {Leggett}, {Levison}, {Luginbuhl},
  {Martini}, {Monet}, {Munn}, {Pier}, {Rhodes}, {Riepe}, {Sell}, {Stone},
  {Vrba}, {Walker}, {Westerhout}, {Brucato}, {Reid}, {Schoening}, {Hartley},
  {Read}, \& {Tritton}}]{2003AJ....125..984M}
{Monet}, D.~G., {Levine}, S.~E., {Canzian}, B., {et~al.} 2003, \aj, 125, 984

\bibitem[{{Natta} \& {Testi}(2001)}]{2001A&A...376L..22N}
{Natta}, A. \& {Testi}, L. 2001, \aap, 376, L22

\bibitem[{{Natta} {et~al.}(2004){Natta}, {Testi}, {Muzerolle}, {Randich},
  {Comer{\'o}n}, \& {Persi}}]{2004A&A...424..603N}
{Natta}, A., {Testi}, L., {Muzerolle}, J., {et~al.} 2004, \aap, 424, 603

\bibitem[{{Neuh{\"a}user} {et~al.}(2002){Neuh{\"a}user}, {Brandner}, {Alves},
  {Joergens}, \& {Comer{\'o}n}}]{2002A&A...384..999N}
{Neuh{\"a}user}, R., {Brandner}, W., {Alves}, J., {Joergens}, V., \&
  {Comer{\'o}n}, F. 2002, \aap, 384, 999

\bibitem[{{Neuh{\"a}user} \& {Comer{\'o}n}(1998)}]{1998Sci...282...83N}
{Neuh{\"a}user}, R. \& {Comer{\'o}n}, F. 1998, Science, 282, 83

\bibitem[{{Neuh{\"a}user} \& {Comer{\'o}n}(1999)}]{1999A&A...350..612N}
{Neuh{\"a}user}, R. \& {Comer{\'o}n}, F. 1999, \aap, 350, 612

\bibitem[{{Rieke} \& {Lebofsky}(1985)}]{1985ApJ...288..618R}
{Rieke}, G.~H. \& {Lebofsky}, M.~J. 1985, \apj, 288, 618

\bibitem[{{Rousset} {et~al.}(2003){Rousset}, {Lacombe}, {Puget}, {Hubin},
  {Gendron}, {Fusco}, {Arsenault}, {Charton}, {Feautrier}, {Gigan}, {Kern},
  {Lagrange}, {Madec}, {Mouillet}, {Rabaud}, {Rabou}, {Stadler}, \&
  {Zins}}]{2003SPIE.4839..140R}
{Rousset}, G., {Lacombe}, F., {Puget}, P., {et~al.} 2003, in Presented at the
  Society of Photo-Optical Instrumentation Engineers (SPIE) Conference, Vol.
  4839, Adaptive Optical System Technologies II. Edited by Wizinowich, Peter
  L.; Bonaccini, Domenico. Proceedings of the SPIE, Volume 4839, pp. 140-149
  (2003)., ed. P.~L. {Wizinowich} \& D.~{Bonaccini}, 140--149

\bibitem[{{Siegler}(2007)}]{2007lyot.confR..45S}
{Siegler}, N. 2007, in Proceedings of the conference In the Spirit of Bernard
  Lyot: The Direct Detection of Planets and Circumstellar Disks in the 21st
  Century. June 04 - 08, 2007. University of California, Berkeley, CA, USA.
  Edited by Paul Kalas., ed. P.~{Kalas}, 45

\bibitem[{{Stassun} {et~al.}(2006){Stassun}, {Mathieu}, \&
  {Valenti}}]{2006Natur.440..311S}
{Stassun}, K.~G., {Mathieu}, R.~D., \& {Valenti}, J.~A. 2006, \nat, 440, 311

\bibitem[{{Stelzer} {et~al.}(2004){Stelzer}, {Micela}, \&
  {Neuh{\"a}user}}]{2004A&A...423.1029S}
{Stelzer}, B., {Micela}, G., \& {Neuh{\"a}user}, R. 2004, \aap, 423, 1029

\end{thebibliography}

\onlfig{6}{
\begin{figure*}
   \resizebox{\hsize}{!}{\includegraphics[angle=90]{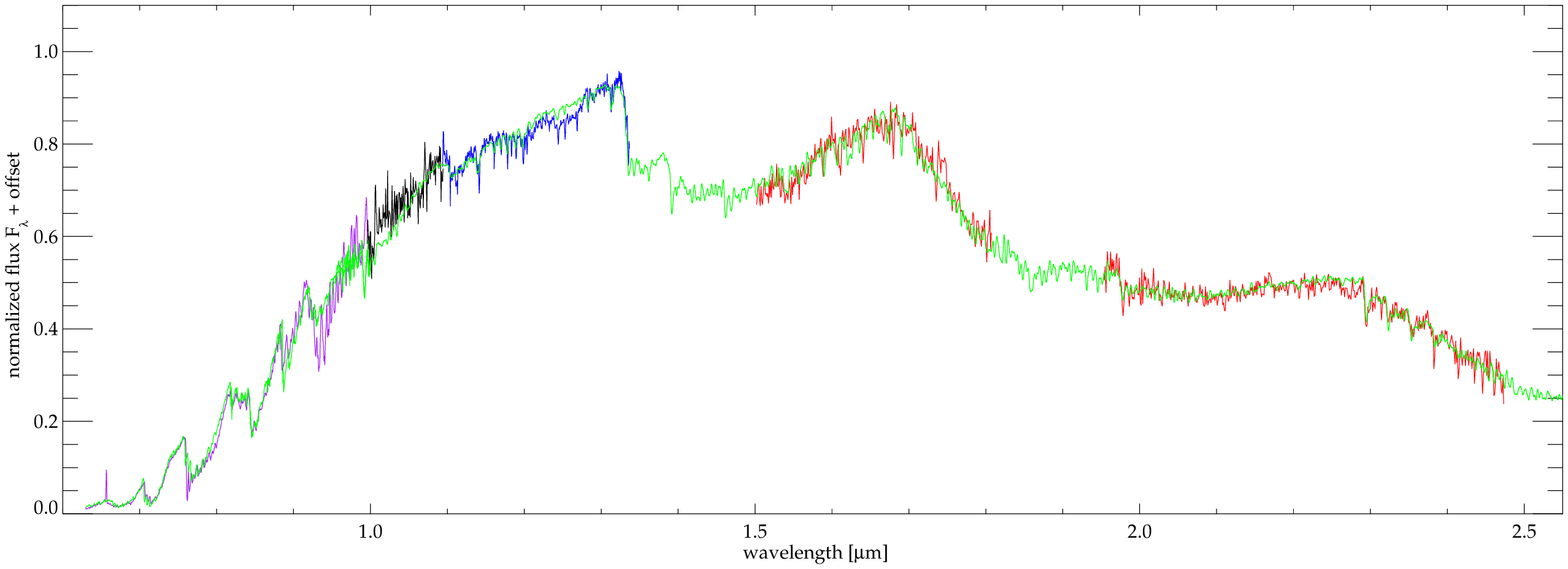}}
   \caption{Spectrum of both components from the optical part to the near-infrared combined from data from \citet{2000A&A...359..269C} (purple), and data from the ESO Paranal and La Silla instruments ISAAC (blue) and SofI (black: blue arm spectra, red: red arm spectra) retrieved from the ESO Science Archive Facility. The best fit to the spectrum, a GAIA model computed with the PHOENIX code by \citet{2005ESASP.576..565B} for a temperature of 3000\,K, $\log{g}$ of 4.0 and a visual extinction of 4.3 is superimposed in green. Note that the missing
flux depression in J-band (between 1.2 and 1.3 micron) as well as in the peak of the H-band (around 1.65 micron) is due to missing FeH opacities in the spectral models. Note the strong H$\alpha$ emission in Cha H$\alpha$\,2 at 0.6568 $\mu$m.}
   \label{SpecAllOnline}
   \end{figure*}
}

\end{document}